\documentstyle[sprocl]{article}

\def\bq{\begin{eqnarray*}}
\def\eq{\end{eqnarray*}}

\begin{document}

\title{ STATUS OF NNLO 3-JET CALCULATIONS\\ }

\author{ STEFAN WEINZIERL }

\address{Max-Planck-Institut f\"ur Physik 
    (Werner-Heisenberg-Institut), F\"ohringer
    Ring 6, D-80805 M\"unchen, Germany}

\maketitle\abstracts{
The process $e^+ e^- \rightarrow \mbox{3 jets}$ offers the opportunity to measure the strong coupling constant. 
For an accurate determination, precise theoretical calculations are necessary. 
I will give an overview on the status of the next-to-next-to-leading order calculations.
}
  
\section{Introduction}
\label{sec:intro}

Today's and future precision measurements in particle physics
require accurate theoretical predictions in order to extract the values of the fundamental
parameters of the theory from experiment.
The precise values of these constants also serve to constrain the parameter space of models
of new physics.
To this aim existing next-to-leading order (NLO) predictions have to be extended
to next-to-next-to-leading order (NNLO).
While for inclusive observables like the total hadronic cross section in
$e^+ e^-$ annihilation this step has been taken long time ago, the situation
is much more complicated for exclusive quantities like event shapes or jet rates.
One prominent process where a NNLO calculation for fully exclusive observables 
is desirable is $ e^+ e^- \rightarrow 3 \; \mbox{jets}$.
The strong coupling constant $\alpha_s$ can be measured by using the
data for $e^+ e^- \rightarrow \mbox{3 jets}$.
A NNLO calculation is expected to reduce significantly
the theoretical uncertainty in the extraction of $\alpha_s$.
A program for a fully exclusive NNLO calculation is flexible and allows to take into account
complicated detector geometries and jet definitions.
The observable may even be defined after the program has been written.
The only requirement on the observable is infrared-safety.
To construct an NNLO program the following ingredients are needed:

- the relevant scattering amplitudes up to two loops;

- a method to cancel infrared divergences;\\
There has been significant progress in the past years for each of these items and I will review
them in the remainder of this talk.

\section{The Calculation Of Two-Loop Amplitudes}
\label{sec:twoloop}

The NNLO calculation of $e^+ e^- \rightarrow \mbox{3 jets}$ requires the following amplitudes:
the Born amplitudes for $e^+ e^- \rightarrow \mbox{5 partons}$, 
the one-loop amplitudes for $e^+ e^- \rightarrow \mbox{4 partons}$
and the two-loop amplitudes for $e^+ e^- \rightarrow \mbox{3 partons}$.
The necessary Born amplitudes \cite{Berends:1989yn,Hagiwara:1989pp}
and the one-loop amplitudes 
\cite{Bern:1997ka,Glover:1997eh}
have been obtained already a while ago.
The most complicated parts are the two-loop amplitudes.
This is due to 
new types of Feynman integrals, corresponding
to Feynman diagrams with two loops and an external off-shell leg.
Methods invented to tackle these integrals comprise the use
of the Mellin-Barnes formula 
\cite{Smirnov:2000vy},
the application of differential equations and integration-by-parts identities
\cite{Gehrmann:1999as}
as well as the use of nested sums
\cite{Moch:2001zr}.
The results are expressed in terms of (multiple) polylogarithms, thus extending the well known
set of basic functions for one-loop integrals (logarithm and dilogarithm) 
and making contact with recent developments in mathematics.
In addition to the results for the basic integrals, various reduction algorithms are also used:
An algorithm by Tarasov 
\cite{Tarasov:1996br}
allows to convert integrals with the loop momentum in the numerator
into scalar integrals with possibly raised powers of the propagators and shifted dimension.
An efficient algorithm by Laporta 
\cite{Laporta:2000dc}
expresses a large set of unknown integrals in terms of a few
``master'' integrals.
With these tools the two-loop amplitudes for $e^+ e^- \rightarrow q g \bar{q}$
have been calculated
\cite{Garland:2001tf,Moch:2002hm}.
The two-loop amplitude has the colour decomposition
\bq
{\cal A}^{(2)}_3 & = & 
 N^2 {\cal A}^{(2)}_{3,1} + {\cal A}^{(2)}_{3,2} + \frac{1}{N^2} {\cal A}^{(2)}_{3,3}
 + N_f N {\cal A}^{(2)}_{3,4} + \frac{N_f}{N} {\cal A}^{(2)}_{3,5} + N_f^2 {\cal A}^{(2)}_{3,6}
 \nonumber \\
 & &
 + {\cal A}^{(2)}_{3,vec} + {\cal A}^{(2)}_{3,ax}.
\eq
${\cal A}^{(2)}_{3,vec}$ and ${\cal A}^{(2)}_{3,ax}$ are the contributions resulting from 
the vector and axial vector coupling of the photon/Z-boson to a closed quark loop. 
The Durham group calculated the partial amplitudes ${\cal A}^{(2)}_{3,1}$ - ${\cal A}^{(2)}_{3,6}$ as well as
${\cal A}^{(2)}_{3,vec}$.
Our group obtained results for ${\cal A}^{(2)}_{3,4}$ - ${\cal A}^{(2)}_{3,6}$.
The results from the two groups for ${\cal A}^{(2)}_{3,4}$ - ${\cal A}^{(2)}_{3,6}$
agree analytically.
The partial amplitude ${\cal A}^{(2)}_{3,ax}$ remains to be calculated, although it is expected 
that this partial amplitude gives a negligible numerical contribution.

\section{Cancellation Of Infrared Divergences}
\label{sec:ir}

At the next-to-next-to-leading order level the ingredients for the third order term in the perturbative
expansion for quantities depending on $n$ resolved ``hard'' partons
are the already mentioned $n$-parton two-loop amplitudes, the $(n+1)$-parton one-loop amplitudes
and the $(n+2)$ Born amplitudes.
Taken separately, each one of these contributions is infrared divergent. Only the sum of all contributions
is infrared finite.
Infrared divergences occur already at next-to-leading order.
At NLO real and virtual corrections contribute.
The virtual corrections contain the loop integrals and can have,
in addition to ultraviolet divergences, infrared divergences.
If loop amplitudes are calculated in dimensional regularisation,
the IR divergences manifest themselves as
explicit poles in the 
dimensional regularisation parameter $\varepsilon=2-D/2$.
These poles cancel with similar poles arising from
amplitudes with additional partons but less internal loops, when integrated over phase space regions where
two (or more) partons become ``close'' to each other.
In general, the Kinoshita-Lee-Nauenberg theorem
\cite{Kinoshita:1962ur,Lee:1964is}
guarantees that any infrared-safe observable, when summed over all 
states degenerate according to some resolution criteria, will be finite.
However, the cancellation occurs only after the integration over the unresolved phase space
has been performed and prevents thus a naive Monte Carlo approach for a fully exclusive
calculation.
It is therefore necessary to cancel first analytically all infrared divergences and to use
Monte Carlo methods only after this step has been performed.
At NLO, general methods to circumvent this problem are known.
This is possible due to the universality of the singular behaviour
of the amplitudes in soft and collinear limits.
Examples are the phase-space slicing method
and the subtraction method.
Within the subtraction method 
\cite{Frixione:1996ms,Catani:1997vz}
one subtracts a suitable approximation term $d\sigma^A$ 
from the real corrections $d\sigma^R$.
This approximation term must have the same singularity structure as the real corrections.
If in addition the approximation term is simple enough, such that it can be integrated analytically
over a one-parton subspace, then the result can be added back to the virtual corrections $d\sigma^V$.
Since by definition $d\sigma^A$ has the same singular behaviour as $d\sigma^R$, 
the combination $(d\sigma^R-d\sigma^A)$ is integrable
and can be evaluated numerically.
Secondly, the analytic integration of $d\sigma^A$ over the one-parton subspace will yield
the explicit poles in $\varepsilon$ needed to cancel the corresponding poles in $d\sigma^V$.
At NNLO this generalizes as follows
\cite{Weinzierl:2003fx}:
\bq
\langle {\cal O} \rangle_n^{NNLO} & = &
 \int 
               {\cal O}_{n+2} \; d\sigma_{n+2}^{(0)} 
             - {\cal O}_{n+1} \circ d\alpha^{(0,1)}_{n+1}
             - {\cal O}_{n} \circ d\alpha^{(0,2)}_{n} 
 \nonumber \\
& &
 + \int 
                 {\cal O}_{n+1} \; d\sigma_{n+1}^{(1)} 
               + {\cal O}_{n+1} \circ d\alpha^{(0,1)}_{n+1}
               - {\cal O}_{n} \circ d\alpha^{(1,1)}_{n}
 \nonumber \\
& & 
 + \int 
                 {\cal O}_{n} \; d\sigma_n^{(2)} 
               + {\cal O}_{n} \circ d\alpha^{(0,2)}_{n}
               + {\cal O}_{n} \circ d\alpha^{(1,1)}_{n}.
 \nonumber
\eq
Here $d\alpha_{n+1}^{(0,1)}$ is a subtraction term for single unresolved configurations
of Born amplitudes.
This term is already known from NLO calculations.
The term $d\alpha_n^{(0,2)}$ is a subtraction term 
for double unresolved configurations.
Finally, $d\alpha_n^{(1,1)}$ is a subtraction term
for single unresolved configurations involving one-loop amplitudes.

To construct these terms the universal factorisation properties of 
QCD amplitudes in unresolved limits are essential.
QCD amplitudes factorise if they are decomposed into primitive
amplitudes.
Primitive amplitudes are defined by
a fixed cyclic ordering of the QCD partons,
a definite routing of the external fermion lines through the diagram
and the particle content circulating in the loop.
One-loop amplitudes factorise in single unresolved limits as
\cite{Bern:1994zx,Bern:1998sc,Kosower:1999xi,Catani:2000pi}
\bq
\label{oneloopfactformula}
A^{(1)}_{n}
  & = &
  \mbox{Sing}^{(0,1)} 
  \cdot A^{(1)}_{n-1} +
  \mbox{Sing}^{(1,1)} \cdot A^{(0)}_{n-1}.
\eq
Tree amplitudes factorise in the double unresolved limits as
\cite{Berends:1989zn,Gehrmann-DeRidder:1998gf,Campbell:1998hg,Catani:1998nv,DelDuca:1999ha}
\bq
\label{factsing}
A^{(0)}_{n}
  & = &
  \mbox{Sing}^{(0,2)} \cdot A^{(0)}_{n-2}.
\eq
The subtraction terms can be derived by working in the axial gauge. In this gauge only diagrams where the emission
occurs from external lines are relevant for the subtraction terms.
Alternatively, they can be obtained from off-shell currents and
antenna factorisation 
\cite{Kosower:1998zr}.

\section{Outlook}
\label{sec:outlook}

In this talk I reviewed the status of NNLO 3-jet calculations.
With the progress we witnessed in the field in the last years we can expect to obtain numerical results rather soon
and to extend
existing numerical programs for NLO predictions
on $e^+ e^- \rightarrow 4 \;\mbox{jets}$ 
\cite{Dixon:1997th,Nagy:1998bb,Campbell:1998nn,Weinzierl:1999yf}
towards NNLO predictions
for $e^+ e^- \rightarrow 3 \;\mbox{jets}$.
In fact, a result for a particular colour structure for the thrust observable 
was announced recently \cite{Gehrmann-DeRidder:2004xe}.


\end{document}